\documentclass[a4paper]{jpconf}

\usepackage{graphicx}
\usepackage{amsbsy}
\usepackage{amsmath}
\usepackage{amsfonts}
\usepackage{amssymb}
\usepackage{epsfig}
\usepackage{sidecap}
\usepackage[dvipsnames]{xcolor}
\usepackage{xspace}
\usepackage{citesort}
\usepackage{wrapfig}
\usepackage{placeins}
\usepackage{arydshln} %dashed lines in tables
\usepackage{empheq}

\newcommand{\He}{\ensuremath{\mathrm{He}}\xspace}
\newcommand{\HeI}{\ensuremath{{\mathrm{He}^+}}\xspace}

\newcommand{\alp}{\ensuremath{{\mathrm{He}^{2+}}}\xspace}
\newcommand{\HH}{\ensuremath{\mathrm{H}}\xspace}
\newcommand{\Hnot}{\ensuremath{\mathrm{H}^0}\xspace}
\newcommand{\HHnot}[1]{\ensuremath{\mathrm{H^0_{(#1)}}}\xspace}
\newcommand{\Henot}{\ensuremath{\mathrm{He}^0}\xspace}
\newcommand{\HHenot}[1]{\ensuremath{\mathrm{He^0_{(#1)}}}\xspace}

\newcommand{\HI}{\ensuremath{{\mathrm{H}^+}}\xspace}
\newcommand{\p}{\ensuremath{\mathrm{p}}\xspace}

\newcommand{\pui}{\ensuremath{\mathrm{PUI}}\xspace}
\newcommand{\Hpui}{\ensuremath{\mathrm{H^+,PUI}}\xspace}

\newcommand{\core}{\ensuremath{\mathrm{core}}\xspace}

\begin{document}
	\title{Magnetohydrodynamic/Kinetic Modeling of the Interaction between the Solar Wind and the Local Interstellar Medium with He$^+$, He$^{2+}$, and Pickup H$^+$ ions and H, He Atoms}
	
	\author{Federico Fraternale}
	
	\address{The University of Alabama in Huntsville, Center for Space Plasma and Aeronomic Research, Huntsville AL, 35899}
	
	\author{Nikolai~V.~Pogorelov}
	\address{Center for Space Plasma and Aeronomic Research, The University of Alabama in Huntsville, Huntsville, AL 35899, USA }
	
	\address{Department of Space Science, The University of Alabama in Huntsville, Huntsville, AL 35899, USA \\}
	
	\author{Ratan~K.~Bera}
	\address{Center for Space Plasma and Aeronomic Research, The University of Alabama in Huntsville, Huntsville, AL 35899, USA }

	\ead{federico.fraternale@uah.edu}
	
	\begin{abstract}
The interaction of the solar wind with the local interstellar medium (LISM) spans a wide range of interacting particle populations, energies, and scales. Sophisticated models are required to capture the global picture, interpret near-Earth observations, and ultimately understand the properties of the LISM at distances of thousands of AUs, where the medium is presumed to be unperturbed by this interaction.
We present a new extension of our MHD-plasma/kinetic-neutral heliospheric model, implemented within the Multi-Scale Fluid-Kinetic Simulation Suite (MS-FLUKSS). The new model treats singly and doubly charged helium ions, pickup protons, and electrons as separate, self-consistently coupled populations, interacting through six charge exchange processes and photoionization with kinetically treated neutral hydrogen and helium atoms. In this paper, we provide detailed information on the implementation, including new fits for the charge-exchange cross sections, and demonstrate the functionality and performance of the new code.
		
	\end{abstract}
	
	\section{Introduction}\label{sec:intro}

The interaction of the solar wind (SW) with the local interstellar medium (LISM), which forms the heliosphere, is deeply influenced by the ionization processes of interstellar neutral atoms (ISN) \cite{blum1969}. Charge exchange, photoionization, electron impact and recombination processes, as well as elastic collisions provide fundamental couplings between ions and neutral atoms, hence between solar and interstellar matter. These processes are ultimately responsible, to a large extent, for the shape of the heliosphere and the properties of the very local interstellar medium (VLISM) perturbed by the heliosphere presence, which may extend to distances of thousands of AUs (e.g., \cite{pogorelov2015,pogorelov2017b,zhang2020}).

A handful of space missions have provided in situ data from the outer heliosphere, including Voyagers 1 and 2 (V1, V2), Ulysses, and New Horizons (NH). Fundamental data also come from neutral atom observations by the Interstellar Boundary Explorer (IBEX), which measures the fluxes of neutral atoms reaching 1 AU from all directions in the sky. Scheduled for launch in 2025, the Interstellar Mapping and Acceleration Probe (IMAP) \cite{mccomas18a} will provide similar observations with higher accuracy and capability to distinguish between particle species. These observations include low-energy interstellar neutrals (ISNs) and energetic neutral atoms (ENAs). The latter are typically produced by neutralization of pickup ions (PUIs) due to charge exchange occurring in various heliospheric regions, including the outer heliosheath (OHS).

Because the energy-dependent fluxes of atoms reaching 1 AU carry an imprint of the properties of heliospheric and interstellar plasma along the whole path from the far LISM to the observation point, they serve as a fundamental tool for understanding the physics of the outer heliosphere. However, interpreting these observations always requires some form of modeling, and still notable discrepancies exist between different models and observations \cite{galli2023}.

Since the velocity distribution functions (VDFs) of  \HH and \He atoms are neither Maxwellian nor isotropic \cite{wood2019,swaczyna2021,fraternale2021b},  3D time-dependent global modeling with a self-consistent, kinetic treatment of neutral atoms is essential for making accurate interpretations and predictions. This is especially true in light of  future IMAP data, which will additionally perform helium ENA measurements. Observations of low-energy helium ISN atoms from Ulysses and IBEX-Lo are particularly compelling due to the \He abundance and their relatively low ionization rates, which result in large collision mean free paths. These observations have been instrumental in inferring the properties of the LISM surrounding us, see, e.g., \cite{lallement2004b,lallement2005,moebius2004,bzowski2012,bzowski2019,swaczyna2023a}. At higher energies, future observations may provide critical insights into the structure of the heliosphere \cite{swaczyna2017b}.

Recently, we have been developing a continuously evolving MHD-plasma/kinetic-neutrals model which self-consistently incorporates helium and hydrogen \cite{fraternale2021b,fraternale2023,fraternale2024a}. In parallel, we have extended our multifluid models which, since \cite{pogorelov2016}, have been using a separate description of pickup protons, including a proper treatment of PUIs crossing the heliospheric termination shock (TS) \cite{bera2023a,bera2024a}, based on hybrid-kinetic modeling. 

In \cite{fraternale2024b}, we have developed the method to solve unsteady (solar cycle) MHD/kinetic problems, which are particularly challenging, albeit unavoidable. In that study, we showed how significantly the solution might depend on the assumptions made about the ion VDF, when modeling charge exchange in different regions of space. On the other hand, we have previously shown the impact of treating electrons  \cite{fraternale2023} and PUIs \cite{bera2023a,bera2024a} as distinct fluids on the global solution. However, PUIs have not been included yet into the MHD/kinetic model as a separate ion population, and helium ions in the SW regions were represented by only one population (either singly or doubly charged).

In this study, we present a new, more comprehensive MHD/kinetic model, particularly suitable for ISN and ENA flux calculations. We extend the model of \cite{fraternale2024b} by including a separate treatment of pickup \HI ions and distinct equations for \HeI and \alp ions, allowing \HeI ions to exist and be produced everywhere. Additionally, we incorporate three additional charge exchange processes that are particularly important for generating secondary and higher-order populations of neutral \He and \HH atoms, including the energetic ones. We present new analytic fits for the charge-exchange cross section data \cite{barnett1990} with small relative errors at all collision energies.  The model is fully 3D and self-consistent in all its components.

In Section \ref{sec:model}, we provide a detailed description of the physical model. In Section \ref{sec:results}, we discuss the performance of the new code and show the source terms obtained from the first two simulations using the updated model. We demonstrate that our software is fully operational and efficient.  Conclusions follow in Section \ref{sec:conclusions}

	\section{Physical model}\label{sec:model}
	Our physical model is implemented in the Multi-Scale Fluid-Kinetic Simulation Suite (MS-FLUKSS,\cite{pogorelov2014}). It consists of an MHD description of plasma and a kinetic description for neutral \HH and \He atoms. 	The plasma description is based on the solution of a set of the conservation laws for the mixture of thermal protons, pickup
	protons, electrons, and helium ions, hereinafter referred to as the plasma mixture, and auxiliary continuity and pressure equations for electrons, \HI PUIs, \HeI and \alp ions. All populations are assumed to comove on the MHD scales, due to the effects of turbulence and Coulomb collisions (in the LISM). Neutral \HH and \He atoms are governed by their respective Boltzmann equations, which account for the production and loss processes and are solved using a Monte Carlo method \cite{malama1991}. The full set of equations is as follows:

\begin{empheq}[left=\empheqlbrace, right=]{align}
	&\partial_t \rho + \boldsymbol{\nabla}\cdot(\rho{\bf u}) = \sum_s {S}^\rho_s, \\
	&\partial_t(\rho {\bf u}) + \boldsymbol{\nabla}\cdot\left[\rho{\bf uu} + (p + \mathbf{B}^2/8\pi){\bf I} - \frac{1}{4\pi}{\bf BB}\right] = \sum_s{\bf S}^\mathrm{m}_s, \\
	&\partial_t E + \boldsymbol{\nabla}\cdot\left[(E + p + \mathbf{B}^2/8\pi){\bf u} - \frac{1}{4\pi}({\bf B \cdot u}){\bf B}\right] = \sum_s{S}^{E}_s, \\
	&\partial_t \mathbf{B} + \nabla \cdot (\mathbf{uB} - \mathbf{Bu}) = 0, \\
		\nonumber \\
	&\partial_t \rho_{\HeI} + \boldsymbol{\nabla}\cdot({\bf u} \, \rho_\HeI) = S^\rho_{\HeI}, \\
	&\partial_t p_\HeI + \boldsymbol{\nabla}\cdot({\bf u} \, p_\HeI) = (1 - \gamma) \, p_\HeI \nabla \cdot {\bf u} + S^p_{\HeI} + {Q^\mathrm{C}_{\HeI}}, \\
	&\partial_t \rho_{\alp} + \boldsymbol{\nabla}\cdot({\bf u} \, \rho_\alp) = S^\rho_{\alp}, \\
	&\partial_t p_\alp + \boldsymbol{\nabla}\cdot({\bf u} \, p_\alp) = (1 - \gamma) \, p_\alp \nabla \cdot {\bf u} + S^p_{\alp} + {Q^\mathrm{C}_{\alp}}, \\
	&\partial_t \rho_{\Hpui} + \boldsymbol{\nabla}\cdot({\bf u} \, \rho_\Hpui) = S^\rho_{\Hpui}, \\
	&\partial_t p_\Hpui + \boldsymbol{\nabla}\cdot({\bf u} \, p_\Hpui) = (1 - \gamma) \, p_\Hpui \nabla \cdot {\bf u} + S^p_{\Hpui} + {Q^\mathrm{C}_{\Hpui}} - {Q^\mathrm{turb}}, \label{eq:PUI-press}\\
	&\partial_t p_\mathrm{e} + \boldsymbol{\nabla}\cdot({\bf u} \, p_\mathrm{e}) = (1 - \gamma) \, p_\mathrm{e} \nabla \cdot {\bf u} + S^p_\mathrm{e} + {Q^\mathrm{C}_\mathrm{e}} + \alpha Q^\mathrm{turb}, \\
	&\partial_t \psi + \mathbf{u} \nabla \psi = 0,  \label{eq:levelset} \\
	\nonumber\\
	&\partial_t f_\HH + \mathbf{v} \cdot \nabla f_\HH = \nonumber\\
	&\sum_s f_s (\mathbf{x}, \mathbf{v}, t) \int f_\HH (\mathbf{x}, \mathbf{v}, t) ~|\mathbf{v}_\HH - \mathbf{v}|~ \sum_p\sigma_{\mathrm{cx},p}(|\mathbf{v}_\HH - \mathbf{v}|) {\rm d}^3 \mathbf{v}_\HH \nonumber \\
	&- f_\HH (\mathbf{x}, \mathbf{v}, t) \left[\beta_{\rm ph,\HH}+ \sum_s \int f_s (\mathbf{x}, \mathbf{v}, t) ~|\mathbf{v} - \mathbf{v}_s| ~\sum_l\sigma_{\mathrm{cx},l}(|\mathbf{v} - \mathbf{v}_s|) {\rm d}^3 \mathbf{v}_s  \right], \label{eq:BoltzmannH}\\
	&\partial_t f_\He + \mathbf{v} \cdot \nabla f_\He + \mathbf{g} \cdot \nabla_v f_\He =\nonumber\\
	 &\sum_s f_s (\mathbf{x}, \mathbf{v}, t) \int f_\He (\mathbf{x}, \mathbf{v}, t) ~|\mathbf{v}_\He - \mathbf{v}|~ ~ \sum_p\sigma_{\mathrm{cx},p}(|\mathbf{v}_\He - \mathbf{v}|) {\rm d}^3 \mathbf{v}_\He \nonumber \\
	&- f_\He (\mathbf{x}, \mathbf{v}, t) \left[ \beta_{\rm ph,\He} + \sum_s \int f_s (\mathbf{x}, \mathbf{v}, t) ~|\mathbf{v} - \mathbf{v}_s|~ \sum_l\sigma_{\mathrm{cx},l}(|\mathbf{v} - \mathbf{v}_s|) {\rm d}^3 \mathbf{v}_s  \right],\label{eq:BoltzmannHe}
\end{empheq}
where
    \begin{equation}\label{eq:sourceP}
		S^p_s=(\gamma-1)\left(S^E_s-{\bf u}\cdot {\bf S}^\mathrm{m}_s+\frac{1}{2}u^2S^\rho_s\right),
	\end{equation}
	\begin{equation}\label{eq:sourceTurb}
		Q^\mathrm{turb}=\frac{1}{2}(\gamma-1) f_\mathrm{D} |{\bf u}_\HH-{\bf u}| V_\mathrm{A}~m_\p S^n_\pui.
	\end{equation} 
    
	\begin{align} \label{eq:CoulombColl}
	Q^\mathrm{C}_s&=k_\mathrm{B} \sum_j^3 n_s \bar\nu_{sj} (T_j-T_s)=\sum_j^3\bar\nu_{sj} \left(\frac{n_s}{n_j}p_j-p_s\right),\\
	\bar\nu_{sj}&=\frac{(\gamma -1) 4\sqrt{2\pi} q^4\Lambda_{sj} n_j\sqrt{m_j m_s}}{k_\mathrm{B}^{3/2}(m_j T_{s}+m_{s}T_j)^{3/2}}.\label{eq:CoulombCollFreq}\\
	%
	%\Lambda_{\e {i}}&=\left\{
     %           \begin{array}{ll}
    %              23-\ln n_\e^\frac{1}{2}T_\e^{-\frac{3}{2}}\ \ \ T_i m_\e/m_i< T_\e < 10~ \mathrm{eV}\\
     %             24-\ln n_\e^\frac{1}{2}T_\e^{-1}\ \ \  T_i m_\e/m_i< 10~\mathrm{eV} < T_\e\\
    %              16-\ln n_i^\frac{1}{2}T_i^{-\frac{3}{2}} \frac{m_{i}}{m_\p}\ \ \ T_\e<T_i m_\e/m_i
     %           \end{array} \right.\label{eq:CoulombLog_ep}\\
%
%  \Lambda_{\p \HeI}	&= 23 -\ln\left[\frac{m_\p+m_\He}{m_\p T_\HeI m_\HeI T_\p} \left(\frac{n_\p}{T_\p}+\frac{n_\HeI}{T_\HeI} \right)^{1/2}  \right].\label{eq:CoulombLog_pHe}
\end{align}

	In addition, we apply to Eq.~\ref{eq:PUI-press} the boundary conditions at the TS developed in \cite{bera2023a,bera2024a} based on extensive hybrid kinetic simulations, to obtain a more realistic pressure of \HI PUIs behind the shock,  which cannot be achieved using the pressure equation alone. A similar approach could be applied in the future for \HeI PUIs as well.
	Here the quantities $\rho_s$, $p_s$, $T_s$, represent the mass density, thermal pressure and temperature of each plasma component $s=\{\HI_\core,\HI_\pui,\HeI,\alp,e^-\}$  (where 'core' denotes the thermal proton population as opposed to the suprathermal PUIs), $\bf{u}$ and $\bf{B}$ are the bulk velocity and magnetic field vectors, respectively, and $u$ and $B$ are their magnitudes.
	The total energy density is $E=\rho (\epsilon + \mathbf{u}^2/2)+\mathbf{B}^2/8\pi$, where $\epsilon$ is the specific internal energy ($\rho \epsilon = p/(\gamma-1)$) and the adiabatic index $\gamma$ is equal to 5/3 for all species. The terms $S_s^{\rho,\mathrm{m},E,p}$ represent the sources in the mass,  momentum, energy, and pressure equations due to charge exchange and photoionization, for each species $s$. These terms result from the solution of the Boltzmann equations, formally given by Eqs.~\ref{eq:BoltzmannH} and \ref{eq:BoltzmannHe}, which are solved using a trajectory splitting Monte Carlo method \cite{malama1991}.  In MHD/Kinetic models, neutral atoms are treated as individual particles moving through the heliosphere. Their trajectories are calculated using the equation of motion, and events like charge exchange occur stochastically, based on local plasma conditions. The integrals on the right-hand side of the Boltzmann equations, which define the source terms for the MHD equations at each plasma cell and constitute the fundamental coupling between neutrals and ions, are obtained by gathering statistics. To achieve sufficient statistical accuracy, particle splitting is employed at specific locations, particularly near grid boundaries, before neutral atoms enter the finer grid level. Additionally, this method allows for the derivation of the VDFs of neutral atoms and their moments as a post-processing step. A detailed description of the method is given in \cite{lipatov2002book}.
    Note that Eqs. \ref{eq:BoltzmannH} and \ref{eq:BoltzmannHe}  are written in a general form, with the right-hand side accounting for the possibility of charge exchange involving mixed species. For hydrogen particles, we currently assume that gravity compensates the radiation pressure force exactly, while the radiation pressure is neglected for helium.  These assumptions have been widely used and have no consequence on the outer heliosphere solution. However, it may be important to implement a separate, time-dependent radiation pressure model, in future versions, in order to use our model to derive more atom VDFs at 1 AU. The terms $Q^\mathrm{C}_s$ represent the energy transfer rate to species $s$ due to Coulomb collisions with other species, with collision frequencies defined by Eq. \ref{eq:CoulombCollFreq}, and Coulomb logarithms as defined in \cite{fraternale2023}. These thermalization terms should be adapted to the case of kappa ion VDFs.
    The $Q^\mathrm{turb}$ term represents the classical energy density rate due to PUI-driven instabilities in the supersonic SW. This term in reality feeds a turbulent cascade, however, we approximate that this energy flux is eventually converted into heat, with 60\% going to protons and 40\% to electrons (because we are not solving turbulence transport equations here). Note that these terms represent the energy transfer between different species and therefore do not appear in the equations for the plasma mixture. The properties of the core protons are derived from the properties of the plasma mixture and those of the remaining species, so no additional equations are solved for them. Ultimately, one advection equation (\ref{eq:levelset}) for the scalar variable $\psi$ is used to track the position of the HP with the Level-set method. A detailed discussion is provided in \cite{fraternale2024b}.%[RKB: SHOULD WE DESCRIBE LEVEL SET $\psi$? IT IS NOT MENTIONED HERE ALTHOUGH WE WRITE AN EQUATION FOR IT. ]
		
	\begin{center}
		\begin{table}[h]
			\caption{\label{tab:processes} Ionization processes in the new model.}
			%\footnotesize\rm
			\centering
			\begin{tabular}{@{}*{3}{l}}
				\br
				Process & Label & Notes \\
				\mr
				\multicolumn{3}{l}{Charge exchange of hydrogen atoms}\\
				\mr
				$ \HI + \Hnot  \to   \Hnot+\HI$  &  (\textit{a}) & Cross sections $\sigma^\HH_\mathrm{cx,a}$, from Eq.~\ref{eq:sigma_A})\\
				$\alp + \Hnot \to \HeI +  \HI  $  &  (\textit{b}) & Cross sections $\sigma^\HH_\mathrm{cx,b}$, from Eq. (\ref{eq:sigma_B})\\
				$\HeI + \Hnot \to  \He + \HI$  &  (\textit{c}) & Cross sections $\sigma^\HH_\mathrm{cx,c}$, from  Eq.~(\ref{eq:sigma_C}) \\
				\mr
				\multicolumn{3}{l}{Charge exchange of helium atoms}\\
				\mr
				$\HeI+\Henot \to \Henot + \HeI$  &  (\textit{d}) & Cross sections $\sigma^\He_\mathrm{cx,d}$, from  Eq.~(\ref{eq:sigma_D})\\
				$\alp + \Henot \to \HeI + \HeI$  &  (\textit{e}) & Cross sections $\sigma^\He_\mathrm{cx,e}$, from  Eq.~(\ref{eq:sigma_E}) \\
				$\alp +  \Henot \to \Henot + \alp $  &  (\textit{f}) & Cross sections $\sigma^\He_\mathrm{cx,f}$, from  Eq.~(\ref{eq:sigma_F})\\ 
				\mr
				\multicolumn{3}{l}{Photoionization}\\
				\mr
				$\Hnot + h\nu \to  \HI + e^-$  &  (\textit{g}) & $\beta_\mathrm{ph}=\beta_\mathrm{ph, 0} (R_0/R)^2$ \\
				$\Henot + h\nu \to  \HeI + e^-$  &  (\textit{h}) &  \\
				\br
			\end{tabular}
		\end{table}
	\end{center}
    
	\subsection{Ionization processes and source terms}
	
In this section, we describe the source terms resulting from various ionization processes, including charge exchange and photoionization. Electron-impact ionization and recombination, as well as  some charge-exchange processes (see, e.g., \cite{grzedzielski2013,swaczyna2017b}) are not yet incorporated into the model. The ionization processes included in the new model, labeled from (\textit{a}) to (\textit{h}), are listed in Table~\ref{tab:processes}, while the details of the processes sorted by region of space are provided in Table~\ref{tab:processes_region}. Hereinafter, as usual, we use subscripts from (0) to (3) to denote neutral atoms originating from corresponding region of space, where (0) represents the distant LISM, (1) the VLISM or outer heliosheath, (2) the inner heliosheath, and (3) the supersonic SW. Note that with this model we choose to treat separately the \HI PUIs born in the supersonic SW. Therefore, in the inner heliosheath (IHS), PUIs are mostly lost, being converted to core protons. However, we include in the PUI population also some PUIs that are injected locally in the inner heliosheath, i.e. those that are  produced when charge exchange occurs between an IHS ion and a fast atom born in the supersonic SW. Moreover, we use the same equations (9-10) to also obtain the bulk properties of \HI PUIs produced in the VLISM. Such PUIs are created when neutral SW atoms from region 3 and 2 reach the outer heliosheath where they experience charge exchange. Our recent study \cite{fraternale2024b} suggests that this may be important in order improve our knowledge of the proton VDF in this region.

	{%
		\renewcommand{\arraystretch}{1.2} % Only affects this table
		\begin{center}
			\begin{table}[h]
				\caption{\label{tab:processes_region} Ionization processes by regions of space. }
				%\footnotesize\rm
				\centering
				%\begin{tabular}{@{}*{2}{l}|@{}*{2}{l}}
				\begin{tabular}{ll|ll}
					%\br
					%    Process & Label & Process & Label \\
					\mr
					\multicolumn{2}{l|}{Region 3 (supersonic SW)} &  \multicolumn{2}{l}{Region 2 (inner heliosheath)}\\
					\mr
					$ \HI_\core + \Hnot_\mathrm{(0,1,2)} \to  \HHnot3 +  \HI_\pui$  &  (a31) & $\ \HI_\core +\Hnot_\mathrm{(0,1,2)} \to  \HHnot2 + \HI_\core $  &  (a21)  \\
					$\HI_\core + \HHnot3 \to \HHnot3 +  \HI_\core $  &  (a32) &  $\  \HI_\core +\HHnot3  \to    \HHnot2+ \HI_\pui$  &  (a22)\\
					$ \HI_\pui +\Hnot_\mathrm{(0,1,2)} \to   \HHnot3 + \HI_\pui$  &  (a33) &  $\  \HI_\pui + \Hnot_\mathrm{(0,1,2)} \to  \HHnot2 +  \HI_\core$  &  (a23)\\
					$\HI_\pui + \HHnot3 \to  \HHnot3 + \HI_\core $  &  (a34) &  $\  \HI_\pui + \HHnot3 \to  \HHnot3 +  \HI_\pui$  &  (a24)\\ \hdashline 
					$ \alp + \Hnot_\mathrm{(0,1,2)}  \to    \HeI +  \HI_\pui $  &  (b31) & $\  \alp + \Hnot_\mathrm{(0,1,2)} \to   \HeI +  \HI_\core $  &  (b21) \\
					$ \alp + \Hnot_\mathrm{(3)}  \to    \HeI+ \HI_\core  $  &  (b32) & $\  \alp + \Hnot_\mathrm{(3)} \to  \HeI +  \HI_\pui$  &  (b22) \\ \hdashline
					$ \HeI + \Hnot_\mathrm{(0,1,2)} \to  \HHenot3 +  \HI_\pui $  &  (c31) & $\  \HeI + \Hnot_\mathrm{(0,1,2)} \to   \HHenot2 + \HI_\core $  &  (c21) \\
					$ \HeI + \Hnot_\mathrm{(3)}  \to   \HHenot3 + \HI_\core $  &  (c32) & $\  \HeI + \Hnot_\mathrm{(3)} \to  \HHenot2 + \HI_\pui $  &  (c22) \\ \hdashline
					$ \HeI + \Henot_\mathrm{(0,1,2,3)} \to  \HHenot3 + \HeI $  &  (d31) & $\  \HeI + \Henot_\mathrm{(0,1,2,3)} \to  \HHenot2 + \HeI $  &  (d21)\\  \hdashline
					$ \alp + \Henot_\mathrm{(0,1,2,3)} \to \HeI + \HeI $  &   (e31) & $\  \alp + \Henot_\mathrm{(0,1,2,3)} \to \HeI + \HeI $  &   (e21) \\   \hdashline
					$  \alp + \Henot_\mathrm{(0,1,2,3)} \to  \HHenot3 + \alp$  &  (f31) &  $\  \alp + \Henot_\mathrm{(0,1,2,3)} \to  \HHenot2 + \alp $  &  (f21)\\ \hdashline
					$\Hnot_\mathrm{(0,1,2)} + h\nu \to  \HI_{\pui} + e^-$  &  (g31) & $\Hnot_\mathrm{(0,1,2)} + h\nu \to  \HI_\core + e^-$ & (g21)\\
					$\Hnot_\mathrm{(3)} + h\nu \to  \HI_{\core} + e^-$  &  (g32) & $\Hnot_\mathrm{(3)} + h\nu \to  \HI_\pui + e^-$ & (g22)\\ \hdashline
					$\Henot_\mathrm{(0,1,2,3)} + h\nu \to  \HeI + e^-$  &  (h31) & $\ \Henot_\mathrm{(0,1,2,3)} + h\nu \to  \HeI + e^-$  &  (h21)  \\
					\mr
					\multicolumn{2}{l|}{Region 1 (outer heliosheath, VLISM)} &  \multicolumn{2}{l}{Region 0 (LISM)}\\
					\mr
					$ \HI_\core + \Hnot_\mathrm{(0,1)}  \to   \HHnot1 + \HI_\core $  &  (a11) & $\  \HI_\core+\Hnot_\mathrm{(0,1)}  \to   \HHnot0 + \HI_\core $  &  (a01)  \\
					$ \HI_\core + \Hnot_\mathrm{(2,3)} \to  \HHnot1 +  \HI_\pui $  &  (a12) &  $\  \HI_\core + \Hnot_\mathrm{(2,3)}  \to   \HHnot0 + \HI_\pui $  &  (a02)\\
					$ \HI_\pui + \Hnot_\mathrm{(0,1)} \to  \HHnot1 +  \HI_\core  $  &  (a13) &  $\  \HI_\pui + \Hnot_\mathrm{(0,1)} \to   \HHnot0 + \HI_\core $  &  (a03)\\
					$ \HI_\pui + \Hnot_\mathrm{(2,3)} \to \HHnot1 +  \HI_\pui $  &  (a04) &  $\  \HI_\pui + \Hnot_\mathrm{(2,3)} \to   \HHnot0 + \HI_\pui $  &  (a04)\\ \hdashline 
					$ \HeI + \Hnot_\mathrm{(0,1)} \to  \HHenot1 +  \HI_\core $  &  (c11) & $\  \HeI + \Hnot_\mathrm{(0,1)} \to   \HHenot0 + \HI_\core $  &  (c01) \\
					$ \HeI + \Hnot_\mathrm{(2,3)} \to  \HHenot1 +  \HI_\pui $  &  (c12) & $\  \HeI + \Hnot_\mathrm{(2,3)} \to   \HHenot0 + \HI_\pui $  &  (c02) \\ \hdashline
					$\HeI + \Henot_\mathrm{(0,1,2,3)}\to \HHenot1 +\HeI $  &  (d11) & $\  \HeI + \Henot_\mathrm{(0,1,2,3)} \to  \HHenot0 + \HeI $  &  (c01)\\  \hdashline
					$\Hnot_\mathrm{(0,1)} + h\nu \to  \HI_\core + e^-$  &  (g11) & $\Hnot_\mathrm{(0,1)} + h\nu \to  \HI_\core + e^-$ & (g01)\\
					$\Hnot_\mathrm{(2,3)} + h\nu \to  \HI_\pui + e^-$  &  (g12) & $\Hnot_\mathrm{(2,3)} + h\nu \to  \HI_\pui + e^-$ & (g02)\\
					\hdashline
					$\Henot_\mathrm{(0,1,2,3)} + h\nu \to  \HeI + e^-$  &  (h11) & $\Henot_\mathrm{(0,1,2,3)} + h\nu \to  \HeI + e^-$  &  (h01) \\
					\br
				\end{tabular}
			\end{table}
		\end{center}
	}
	
%The ionization source terms 	
%	\begin{align}\label{eq:sources}
%		S^\rho_\alp &= S^\rho_{\alp,\rm b} + S^\rho_{\alp,\rm e}  \qquad (<0),\\
%		S^{\rm m,e,p}_\alp &= S^{\rm m,e,p}_{\alp,\rm b} + S^{\rm m,e,p}_{\alp,\rm e} +S^{\rm m,e,p}_{\alp,\rm f}  ,\\
%		%
%		S^\rho_\HeI &= S^\rho_{\HeI,\rm b} + S^\rho_{\HeI,\rm c} + S^\rho_{\HeI,\rm e} + S\rho_{\HeI, \rm h},\\
%		S^{\rm m,e,p}_\HeI &= S^{\rm m,e,p}_{\HeI,\rm b} + S^{\rm m,e,p}_{\HeI,\rm c} + S^{\rm m,e,p}_{\HeI,\rm d} + S^{\rm m,e,p}_{\HeI,\rm e} +  S^{\rm m,e,p}_{\HeI,\rm h},\\
%		%
%		S^\rho_\Hpui &= S^\rho_{\Hpui ,\rm a} + S^\rho_{\Hpui ,\rm b} + S^\rho_{\Hpui ,\rm c} + S^\rho_{\Hpui ,\rm g}\\
%		S^\rho_{\HI,\core} &= S^\rho_{\HI,\core ,\rm a} + S^\rho_{\HI,\core ,\rm b}  + S^\rho_{\HI,\core,\rm c} + S^\rho_{\HI,\core ,\rm g}\\
%		%
%		S^\rho_\e &= S^\rho_{\e ,\rm g} + S^\rho_{\e ,\rm h} ,\\
%		S^{\rm m,e,p}_\e &= S^{\rm m,e,p}_{\e,\rm g} + S^{\rm m,e,p}_{\e,\rm h}
%	\end{align}

\begin{figure}[t]
	\centering
	\includegraphics[width=0.9\textwidth]{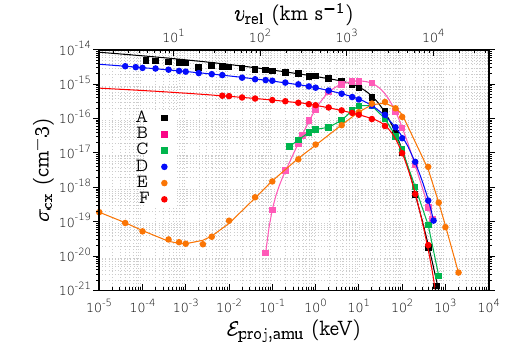}\vskip-10pt
	\caption{\small Charge-exchange cross sections for the processes listed in Table~\ref{tab:processes} are shown as functions of the incident ion (projectile) energy per atomic mass unit, expressed in the target frame (where a neutral atom is at rest). Colored points represent the data from 
    \cite{barnett1990}, while solid lines represent our analytical fits from (\ref{eq:sigma_A})-(\ref{eq:sigma_F}). \label{fig:sigmaCX}}  
\end{figure}
	
We use the charge-exchange cross-section data from Barnett et al. \cite{barnett1990}. Since our model is designed to derive the properties of neutral atoms, including ENAs, we have derived new analytical approximations based on high order logarithmic polynomial fits (Eqs. \ref{eq:sigma_A}-\ref{eq:sigma_F}). These formulae reproduce the data with small relative discrepancies across the entire available energy range (for details on the uncertainties in the data, see \cite{barnett1990}). For the hydrogen process (\textit{a}), we retain the formula used in \cite{fraternale2024b} based on Swaczyna et al. \cite{swaczyna2019a} (S19) and Lindsay \& Stebbings \cite{lindsay2005} (LS05) but introduce a new fit (Eq.~\ref{eq:sigma_A}) for the high-energy range from 0.45 to 220 keV. This range is critical for ENAs, and previous formulas overestimate the cross-section data by up to 50\%. The new fit achieves a maximum relative error below 2\% from 0.45 to 60 keV and below 6\% at higher energies.

The maximum error for the cross sections described by Eqs.(\ref{eq:sigma_B})-(\ref{eq:sigma_D})  is below 5\% across the entire energy range. For Eq. (\ref{eq:sigma_E}), the maximum discrepancy reaches 25\%, with an average error of 6.2\%. Expressions (\ref{eq:sigma_D}) and (\ref{eq:sigma_F}) were previously employed in \cite{fraternale2024b}. The cross-section formulas are provided below and are shown in Fig.~\ref{fig:sigmaCX} alongside the Barnett et al data.

%F=0.2-0.4
%
%B 0.005
%C, D, E  0.01
%F 0.35
	
	\begin{align}\label{eq:sigma_A}
%		&\sigma_a\equiv\sigma_{cx}({\Hnot + \HI}\to \HI+\Hnot)=(4.049-0.447 \log~\mathcal{E})^{2} \left[1-\exp\left(-\frac{60.5}{\mathcal{E}}\right)\right]^{4.5}\times10^{-16}\ \ \mathrm{(cm^2)}.
		&\sigma_a\equiv\sigma_{cx}({ \HI+\Hnot }\to \Hnot+ \HI)=\\ \nonumber&\begin{cases}	
			(4.049-0.447 \log~\mathcal{E})^{2} \left[1-\exp\left(-\frac{60.5}{\mathcal{E}}\right)\right]^{4.5}\times10^{-16}\ \ \mathrm{(cm^2)} \text{~~~for~~} \mathcal{E}\leq 0.45  \text{~keV}\quad  \text{(S19)}\nonumber \\ ~\\ 
\exp\left[\sum_{j=0}^5 c_j (\log \mathcal{E})^j\right] \times 10^4 \, \mathrm{(cm^2)}, \text{~~~for~~} \mathcal{E}> 0.45  \text{~keV}\quad \text{and}\quad \mathcal{E}\leq217  \text{~keV,} \nonumber \\
\mathbf{c} =\big[-4.323635 \times 10^{1}, \, -2.135067 \times 10^{-1}, \, -7.873668 \times 10^{-3},\\
 \qquad \, 2.285425 \times 10^{-2},\, -1.86265\times 10^{-2},\, 1.146968\times 10^{-3}\big], \\ \\
 (4.15-0.531 \log~\mathcal{E})^{2} \left[1-\exp\left(-\frac{67.3}{\mathcal{E}}\right)\right]^{4.5}\times10^{-16}\ \ \mathrm{(cm^2)} \text{~~~for~~} \mathcal{E}> 217  \text{~keV }~~~\text{(LS05).}
		\end{cases}
	\end{align}

%	\begin{align}\label{eq:sigma_B}
%		&\sigma_b\equiv\sigma_\mathrm{cx}({\Hnot+\alp}\to{\HI+\HeI})= \nonumber \\ &\exp\left[\sum_{j=0}^8 c_j (\log \mathcal{E}_\alp)^j\right]\times10^{4}\quad \mathrm{(cm^2)} \quad\text{for}\quad 0.33\leq\mathcal{E}_\alp\leq 2000 \text{~keV,}\\ \nonumber 
%		&\begin{cases}	
%			&c_0 = -4.897315 \times 10^{1} \\
%			&c_1 =  2.670824 \\
%			&c_2 = -1.966020 \times 10^{-2} \\
%			&c_3 =  3.098327 \times 10^{-1} \\
%			&c_4 = -4.576208 \times 10^{-1} \\
%			&c_5 =  1.906637 \times 10^{-1} \\
%			&c_6 = -3.662854 \times 10^{-2} \\
%			&c_7 =  3.350837 \times 10^{-3} \\
%			&c_8 = -1.182081 \times 10^{-4} 
%		\end{cases}
%	\end{align}
\begin{align}\label{eq:sigma_B}
	\sigma_b &\equiv \sigma_\mathrm{cx}({\alp+\Hnot} \to {\HeI+\HI}) = 
	\exp\left[\sum_{j=0}^8 c_j (\log \mathcal{E})^j\right] \times 10^4 \, \mathrm{(cm^2)},  \\
	&\quad \text{for} \quad 0.33 \leq \mathcal{E} \leq 2000 \, \text{keV,} \nonumber\\
	\text{where} \nonumber \\
	\mathbf{c} &= 
	 \big[-4.897315 \times 10^{1}, \, 2.670824, \, -1.966020 \times 10^{-2}, \, 3.098327 \times 10^{-1}, \nonumber \\
	 &\quad -4.576208 \times 10^{-1}, \, 1.906637 \times 10^{-1}, \, -3.662854 \times 10^{-2}, \, 3.350837 \times 10^{-3}, \, -1.182081 \times 10^{-4}\big].\nonumber
\end{align}

\begin{align}\label{eq:sigma_C}
	\sigma_c &\equiv \sigma_\mathrm{cx}({\HeI+\Hnot} \to \Henot+\HI) = 
	\exp\left[\sum_{j=0}^{10} c_j (\log \mathcal{E})^j\right] \times 10^4 \, \mathrm{(cm^2)},  \\
	&\quad \text{for} \quad 1 \leq \mathcal{E} \leq 2800 \, \text{keV,} \nonumber \\
	\text{where} \nonumber \\
	\mathbf{c} &= 
    \big[-4.792325 \times 10^{1}, \, 1.765446, \, -4.318993, \, 8.656753, \nonumber \\
&\quad -8.675877, \, 4.737785, \, -1.510772, \, 2.899103 \times 10^{-1}, \nonumber \\
&\quad -3.303044 \times 10^{-2}, \, 2.060760 \times 10^{-3}, \, -5.426701 \times 10^{-5}\big]\nonumber
\end{align}

	\begin{align}\label{eq:sigma_D}
		&\sigma_d\equiv\sigma_\mathrm{cx}({ \HeI+\Henot }\to{ \Henot+\HeI })=\\ \nonumber&\begin{cases}	
			(3.24-0.288 \log~\mathcal{E})^{2}\times10^{-16}\ \ \mathrm{(cm^2)} \text{~~~for~~} \mathcal{E}\leq 2.36 \text{~keV,}\\
			(5.07-0.707 \log~\mathcal{E})^{1.47}\left[1-\exp\left(-\frac{153}{\mathcal{E}}\right)\right]^{0.824}\times10^{-16}\ \ \mathrm{(cm^2)}\text{~~~for~~} 2.36<\mathcal{E}\leq528 \text{~keV}\\
			(3.97-0.481 \log~\mathcal{E})^{2.35}\left[1-\exp\left(-\frac{269}{\mathcal{E}}\right)\right]^{1.845}\times10^{-16}\ \ \mathrm{(cm^2)}\text{~~~for~~} 528<\mathcal{E}<2100 \text{~keV.}			   
		\end{cases}
	\end{align}

\begin{align}\label{eq:sigma_E}
	\sigma_e &\equiv \sigma_\mathrm{cx}({\alp+\Henot} \to {\HeI + \HeI}) = 
	\exp\left[\sum_{j=0}^{11} c_j (\log \mathcal{E})^j\right] \times 10^4 \, \mathrm{(cm^2)},  \\
	&\quad \text{for} \quad 1.6 \times 10^{-6} \leq \mathcal{E} \leq 8000 \, \text{keV,} \nonumber \\
	\text{where} \nonumber \\
	\mathbf{c} &= 
	% \big[-4.919675 \times 10^{1}, \, 1.051912, \, -7.242226 \times 10^{-2}, \, 1.173159 \times 10^{-2}, 5.116034 \times 10^{-3}\nonumber \\
	% &\quad -8.174245 \times 10^{-4}, \, -1.582438 \times 10^{-4}, \, 6.236905 \times 10^{-6}, \, 1.6978575 \times 10^{-6}, \, 1.463456 \times 10^{-8},\nonumber \\ &\quad-6.354768\times 10^{-9}, 2.241880\times 10^{-10}\big].
   \big[-4.919675 \times 10^{1}, \, 1.051912, \, -7.242226 \times 10^{-2}, \nonumber \\
&\quad 1.173159 \times 10^{-2}, \, 5.116034 \times 10^{-3}, \, -8.174245 \times 10^{-4}, \nonumber \\
&\quad -1.582438 \times 10^{-4}, \, 6.236905 \times 10^{-6}, \, 1.697857 \times 10^{-6}, \nonumber \\
&\quad 1.463456 \times 10^{-8}, \, -6.354768 \times 10^{-9}, \, 2.241880 \times 10^{-10}\big]. \nonumber
\end{align}
%    -0.000000000224188
%	-0.000000006354768
%	 0.000000014634560
%     0.000001697857571
%     0.000006236904907
%	-0.000158243801336
%	-0.000817424555538
%	0.005116034149591
%	0.011731593279685
%	-0.072422260367716
%	1.051912027025654
%	-49.196755425989757
	
	\begin{align}\label{eq:sigma_F}
		&\sigma_f\equiv\sigma_\mathrm{cx}({\alp+\Henot }\to{ \Henot+\alp })=\\ \nonumber&\begin{cases}	
			(2.429-0.255 \log~\mathcal{E})^{1.263}\times10^{-16}\ \ \mathrm{(cm^2)} \text{~~~for~~} \mathcal{E}\leq 1.54 \text{~keV,}\\
			(2.153-0.263\log~\mathcal{E})^{1.523}\left[1-\exp\left(-\frac{568}{\mathcal{E}}\right)\right]^{5.32}\times10^{-16}\ \ \mathrm{(cm^2)}\text{~~~for~~} 1.54<\mathcal{E}<800 \text{~keV.}   
		\end{cases}
	\end{align}

	In all the above expressions, $\mathcal{E}$ is the collision energy expressed in the target frame in keV, i.e. it is the ion energy in the frame where the neutral atom is the target at rest. As in the previous version of our code, we have two separate modules for \HH and \He particles, respectively. Therefore, processes (\textit{a}),(\textit{b}),(\textit{c}), and (\textit{g}) are included in the hydrogen module, while processes (\textit{d}),(\textit{e}), (\textit{f}), and (\textit{h}) are handled by the helium module. As shown in \cite{swaczyna2017b}, process (\textit{c}) may significantly contribute to the conversion of \HeI ions and hydrogen into protons and energetic \He atoms. To account for this, it was necessary to transfer He atoms produced in the hydrogen module to the helium module.
	
	It should be noted that due to the presence of processes (\textit{b}), (\textit{c}), and (\textit{e}), the source term in the continuity equation for the plasma mixture due to charge exchange is no longer zero in this model. %[WHAT IS THE NET SOURCE OF MASS TO THE MIXTURE? YOU ARE SAYING THAT THE SOURCE TERM IN THE CONTINUITY EQUATION FOR THE PLASMA MIXTURE IS NOT ZERO. BESIDES, YOU FREQUENTLY USE THE WORD MIXTURE, WHICH IS CONFUSING, BECAUSE THE MIXTURE ALSO INVOLVES ATOMS.]
	
	Pickup \HeI ions are mostly produced by photoionization (\textit{h}), but not exclusively. For instance, process (\textit{d}) in the supersonic SW typically generates both an energetic (pickup) \HeI ion and a core \HeI ion. In this model, we do not separately track pickup \HeI and \alp ions. This is the reason why processes (\textit{b}), (\textit{d}), (\textit{e}), (\textit{f}), and (\textit{h}) are not split into multiple reactions in Tab.~\ref{tab:processes_region}. The use of kappa VDFs allows us to take into account for the presence of energetic ion tails when the probability of charge exchange is evaluated.

	The probability of a charge exchange event, and the properties of the newborn ions and atoms, are calculated with the following procedure.  For an individual atom $j$, we calculate the integrals
		\begin{gather}
		\mathcal{I}_\beta({\bf v}_{\mathrm{N},j};v_{\rm th,i};\kappa) \equiv \langle \sigma_\mathrm{cx}v_\mathrm{rel}\rangle = \iiint \sigma_{\rm cx}(|{\bf v}_\mathrm{i} - {\bf v}_{\mathrm{N},j} |) |{\bf v}_\mathrm{i} - {\bf v}_{\mathrm{N},j} | f_\mathrm{i}({\bf v}_\mathrm{i}; v_{\rm th,i};\kappa)\, {\rm d} {\bf v}_\mathrm{i}; \label{eq:Ibeta0} \\
		\tilde v_{\rm i} = v_{\rm i}/v_{\rm th,i}, \qquad \tilde v_{\mathrm{N}, j} = v_{\mathrm{N}, j} /v_{\rm th,i}, \qquad \tilde v_{\rm rel}= \sqrt{\tilde v_\mathrm{i}^2+\tilde v_{{\rm N},j}^2-2 \tilde v_\mathrm{i} {\tilde v}_{{\rm N},j}\cos\phi};\\
		\mathcal{I}_\beta({\bf v}_{\mathrm{N},j};v_{\rm th,i};\kappa)  =\int_0^\pi\int_0^{\tilde v_{\rm i,max}} 2\pi \sin(\phi) \tilde v^2_\mathrm{i} \sigma_{\rm ex}(\tilde v_{\rm rel}) \tilde v_{\rm rel} f_{\rm i}(\tilde v_{\rm i};v_{\rm th , i};\kappa) \,{\rm d} {\tilde v}_\mathrm{i}{\rm d} {\phi};\label{eq:Ibeta}\\
		\mathcal{I}_v(v_{\rm th,i}; \tilde v_{\mathrm{N}, j},\tilde v_{\rm i}) = \frac{\int_0^{\tilde v_{\rm i}} \int_0^\pi \sigma(v_{\rm rel}) \sin(\phi) v_{\rm rel}\tilde v_{\rm i}^2 f_{\rm i}(\tilde v_{\rm i})  {\rm d} {\phi} {\rm d} {\tilde v}_\mathrm{i} }{
			\int_0^{\tilde v_{\rm i,max}} \int_0^\pi \sigma(v_{\rm rel}) \sin(\phi) v_{\rm rel}\tilde v_{\rm i}^2 f_{\rm i}(\tilde v_{\rm i}){\rm d} {\phi} {\rm d} {\tilde v}_\mathrm{i}};\label{eq:Iv}\\
		\mathcal{I}_\phi(v_{\rm th,i}; \tilde v_{\mathrm{N}, j},\tilde v_{\rm i},\phi) = \frac{\int_0^\phi   \sigma(v_{\rm rel}) v_{\rm rel} \sin(\phi) \tilde v_{\rm i}^2 f_{\rm i}(\tilde v_{\rm i})      {\rm d} {\phi}}{\int_0^\pi   \sigma(v_{\rm rel}) v_{\rm rel} \sin(\phi) \tilde v_{\rm i}^2 f_{\rm i}(\tilde v_{\rm i})      {\rm d} {\phi}},\label{eq:Iphi}
	\end{gather}
where expressions (\ref{eq:Ibeta})-(\ref{eq:Iphi}) are derived under the assumption for isotropy in the 3D velocity distribution function (VDF) of ions and $f_i$ is the VDF of the $i$-th species normalized by its number density which, in principle, can have any shape but in the current code is either Maxwellian or Lorentzian with parameter $\kappa$.  Here the ion and neutral atom velocities (${\bf v}_\mathrm{i}$ and ${\bf v}_{\mathrm{N},j}$, respectively)  are defined in the plasma reference frame, with the variable $\phi$ representing the angle between the velocity vectors of the colliding neutral atom and the ion. Exactly as in the previous versions of the code, the cumulative integrals (\ref{eq:Ibeta})-(\ref{eq:Iphi}) are pre-calculated and used to draw the velocity and the angle $\phi$ of the newborn ion and atom when a charge exchange event occurs (the azimuthal angle is drawn from a uniform random distribution). They are stored in the code as lookup arrays.
	
In the new code, integrals  (\ref{eq:Ibeta})-(\ref{eq:Iphi}) are calculated, using the individual routines, for each of the processes of Tab. \ref{tab:processes}.  	The frequency at which a neutral atom undergoes charge exchange for each individual process is defined as
\begin{align}
		\beta_{\mathrm{cx},\HH} &= \beta_{\mathrm{cx}, a-\core}+\beta_{\mathrm{cx}, a-\pui}+ \beta_{\mathrm{cx}, b} + \beta_{\mathrm{cx}, c}; \\
		\beta_{\mathrm{cx},\He} &= \beta_{\mathrm{cx}, d}+\beta_{\mathrm{cx}, e}+\beta_{\mathrm{cx}, f};  
	\end{align}
where
	\begin{align}
	\beta_{\mathrm{cx},\text{a-core}} &= \mathcal{I}_{\beta,\text{a-core}}\, n_{\HI_\core}; \\
	\beta_{\mathrm{cx}, \text{a-PUI} } &= \mathcal{I}_{\beta,\text{a-PUI} }\, n_{\Hpui}; \\
	\beta_{\mathrm{cx}, b} &= \mathcal{I}_{\beta,b}\, n_\alp;\\ 
	\beta_{\mathrm{cx}, c} &= \mathcal{I}_{\beta,c}\, n_\HeI;\\ 
	\beta_{\mathrm{cx}, d} &= \mathcal{I}_{\beta,d}\, n_\HeI;\\ 
	\beta_{\mathrm{cx}, e} &= \mathcal{I}_{\beta,e}\, n_\alp;\\ 
	\beta_{\mathrm{cx}, f} &= \mathcal{I}_{\beta,f}\, n_\alp.\\ 
\end{align}

These rates are functions of the local ion temperature, density, and the atom speed. The probability for a neutral \HH or \He atom to experience charge exchange is then obtained as \(\mathcal{P}_{\rm cx,\HH} = \beta_{\rm cx,\HH}\, \delta t \) and \(\mathcal{P}_{\rm cx,\He} = \beta_{\rm cx,\He}\, \delta t \), respectively, where $\delta t$ is a particle time step, which must be smaller than the typical charge exchange time for an accurate description.
	
At each particle time step, and for all particles, we evaluate the probability \(\mathcal{P}\), and determine if charge exchange occurs by checking if \(\mathcal{P} > r\), where \(r \sim \text{U}[0, 1]\) is a random number drawn from a uniform distribution.
	Then, to determine which of the processes occurs, we evaluate the conditioned probabilities. For hydrogen, \( \mathcal{P}_{j} = \beta_{\mathrm{cx}, j} / \beta_{\mathrm{cx}, \HH} \), where \( j \in \{ a\text{-core}, a\text{-PUI}, b, c\} \); for helium, \( \mathcal{P}_{j} = \beta_{\mathrm{cx}, j} / \beta_{\mathrm{cx}, \He} \), where \( j \in \{d, e, f\} \). Now, drawing a second random number \( q \), we determine the type of charge exchange event as follows:
	\begin{itemize}
		\item The \HH event is of type \( a\text{-core} \) if \( q \leq \mathcal{P}_{a\text{-core}} \);
		\item The \HH event is of type \( a\text{-PUI} \) if \( \mathcal{P}_{a\text{-core}} < q \leq \mathcal{P}_{a\text{-core}} + \mathcal{P}_{a\text{-PUI} } \);
		\item The \HH event is of type \( b \) if \( \mathcal{P}_{a\text{-core}} + \mathcal{P}_{a\text{-PUI} } < q \leq \mathcal{P}_{a\text{-core}} + \mathcal{P}_{a\text{-PUI} }+ \mathcal{P}_{b}\);
		\item The \HH event is of type \( c \) otherwise;
		\item The \He event is of type \( d \) if \( q \leq \mathcal{P}_{d} \). 
		\item The \He event is of type \( e \) if \( \mathcal{P}_{d} < q \leq \mathcal{P}_{d} + \mathcal{P}_{e}\). 
		\item The \He event is of type \( f \) otherwise.     
	\end{itemize}
	
	As in the previous models, we have implemented two types of plasma velocity distribution functions (VDF) for each process: the Maxwellian and the Lorentzian (kappa) models. The kappa index is an additional parameter that can differ for each ion species and can vary across different heliospheric regions.

	\section{Results: Test simulations}\label{sec:results}

	We have tested the new model by running a steady-state simulation of the SW-LISM interaction with spherically-symmetric boundary conditions at 1 AU, using the same SW and LISM parameters as in simulation \texttt{B} of \cite{fraternale2024b}, and the boundary conditions for PUIs at the TS. 
	
	To compare the performance, we used Frontera at the Texas Advanced Computing Center (TACC) to perform two simulations on the same grid and computational nodes,  with the old and the new model, respectively. We have used a simulation domain of 1680 au cubed, a Cartesian grid with base size 10 au cubed, and three additional levels of refinement (nested parallelepiped grids with the finest size of 1.25 au cubed). 
	
	As described in detail in the appendix of \cite{fraternale2024b}, the code consists of a plasma stage and a kinetic stage. In the plasma stage, we observe a 40\% increase in wall-time per step due to the inclusion of four additional equations for the density and pressure of \alp ions and PUIs (the number of solved equations increased by 30\%). In the kinetic module, the new code requires the transfer from the plasma module of four additional variables, which are sent to each MPI task via MPI\_Bcast communications. This represents a 40\% increase in the number of variables, resulting in the same increase in MPI communication time and memory requirements. Nevertheless, despite the inclusion of several additional ionization processes, the wall time for the H and He source term calculations increased by only $\sim13$\% and $\sim5$\%, respectively. It is worth noting that, among the various wall-time components, this is the one that primarily determines the overall speed of the code, particularly in time-dependent simulations as discussed in \cite{fraternale2024b}. This is summarized in Tab.~\ref{tab:performance}.
	
\begin{figure}[t]
		\centering\vspace{-50pt}
		\includegraphics[width=0.80\textwidth]{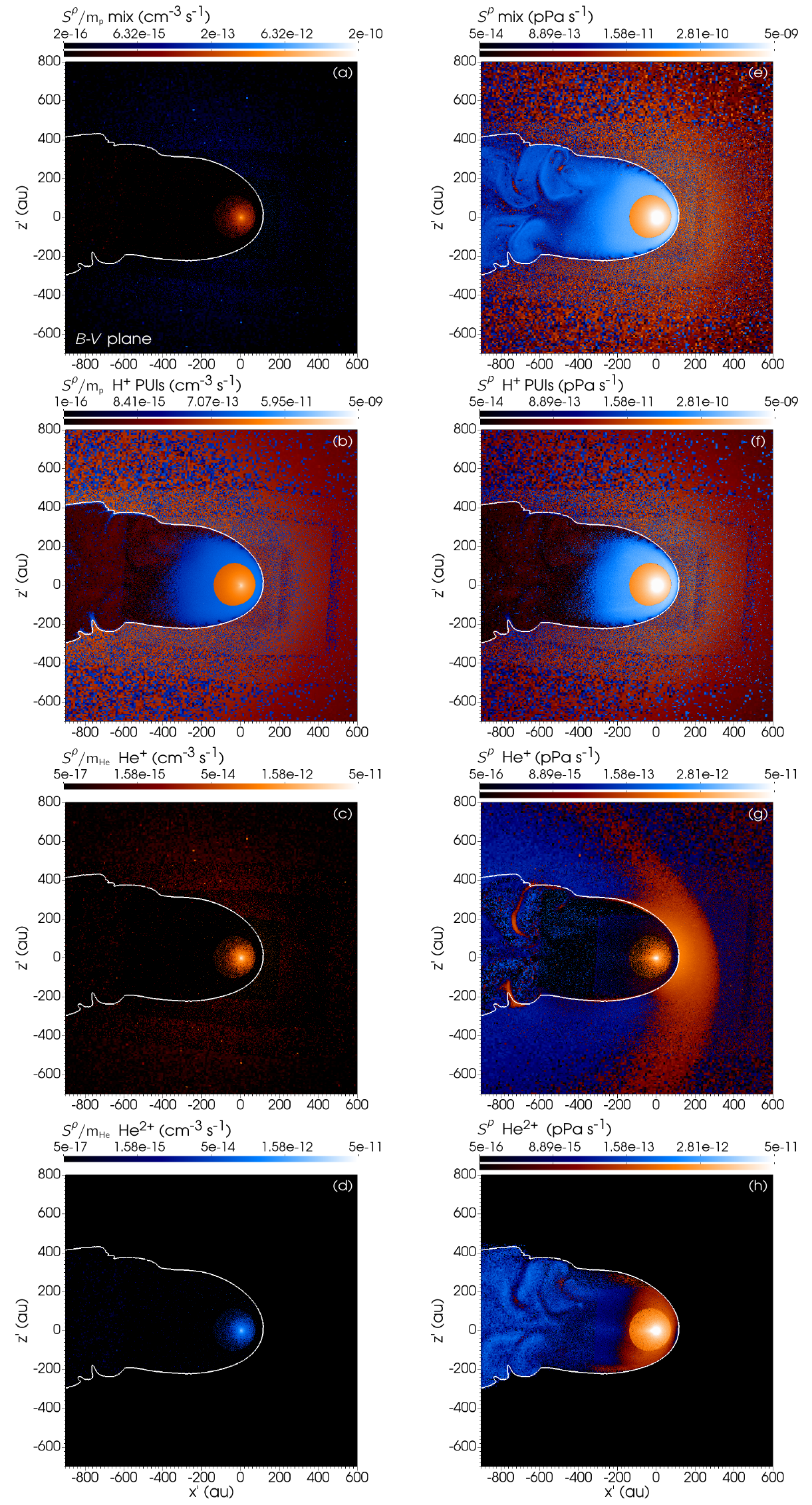}\vskip-15pt
		\caption{\small Simulation \texttt{A}. Visualization in the $B$-$V$ plane of the density (left panels) and pressure (right panels) production (orange color map) and loss (blue) terms  calculated from the new model for each ion species, without the contribution of photoionization (only processes a-f included). The  boundary conditions at 1 AU and the LISM parameters are as in simulation \texttt{B} of \cite{fraternale2024b}. The white curve shows the position of the HP, as tracked with the Level-set method. \label{fig:sources_NoPhi}}  
	\end{figure}

	We demonstrate that the code is operational and efficient by presenting the source terms for density and pressure for all individual plasma species and for the entire plasma mixture. We first ran a simulation (case \texttt{A}) excluding the contribution of the photoionization processes (f,g). The results are shown in Fig.~\ref{fig:sources_NoPhi}. The right column of Fig.~\ref{fig:sources_NoPhi} shows the production (or loss) per unit time in number density for \HI PUI, \alp ions, \HeI ions and the source in density (nucleons) of the ion mixture (from top to bottom). The right panels instead show the production/loss rates in pressure. 
	
\begin{figure}[t]
		\centering\vspace{-50pt}
		\includegraphics[width=0.80\textwidth]{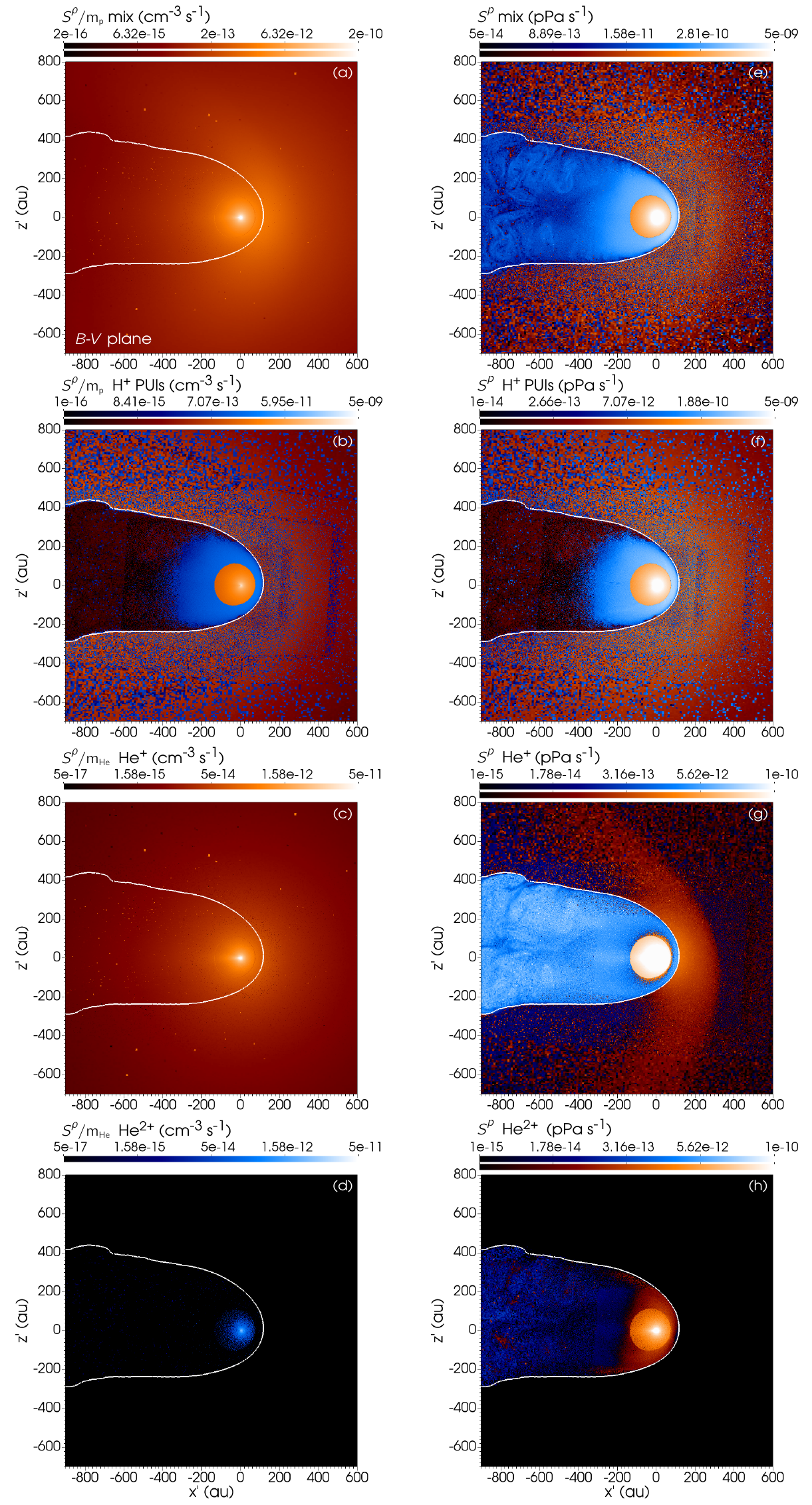}\vskip-15pt
		\caption{\small Simulation \texttt{B}. Density (left panels) and pressure (right panels) production/losses obtained in presence of both charge exchange and photoionization (full model).\label{fig:sources_Phi}  }  
	\end{figure}

	We then ran the second simulation (case \texttt{B}) with the complete model that also includes photoionization processes. The results for this case are shown in Fig.~\ref{fig:sources_Phi}, for the same quantities as in Fig.~\ref{fig:sources_NoPhi}. For both simulations, the kinetic modules ran for 800 years (physical time) per kinetic step to ensure good statistics. 

\begin{figure}[t]
		\centering\vspace{-50pt}
		\includegraphics[width=0.80\textwidth]{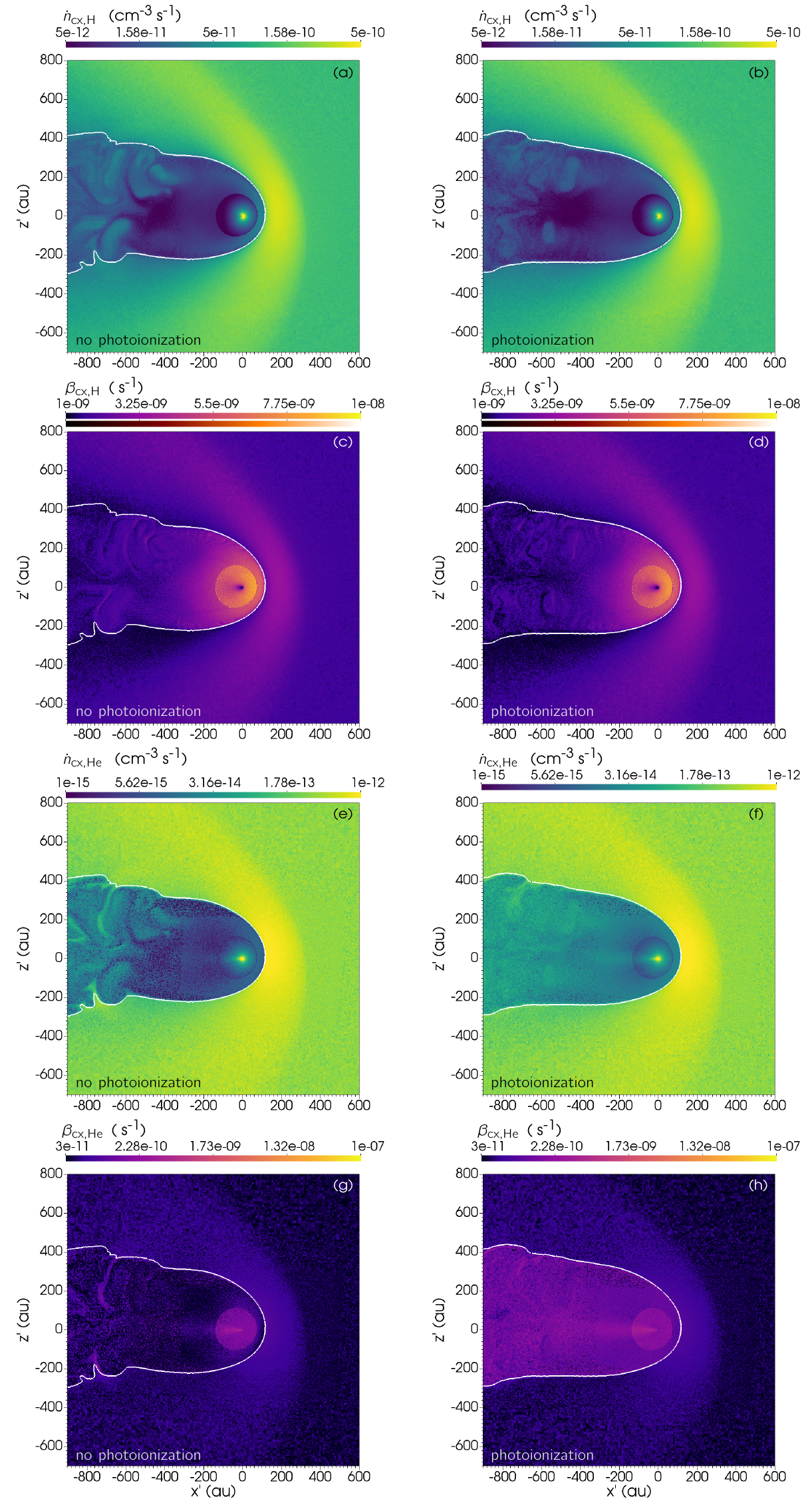}\vskip-5pt
		\caption{\small Number of charge exchange events per unit time and volume, $\dot n_\mathrm{cx}$ (panels a, b, e, f) and charge exchange rates $\beta_\mathrm{cx}$ (panels c, d, g, h) for hydrogen and helium. Simulations \texttt{A} and \texttt{B} are shown in the left and right panels, respectively\label{fig:sources_beta}.  }  
	\end{figure}

    Figure~\ref{fig:sources_beta} instead compares, for the two cases, the total number of charge exchange events per unit time and volume ($\dot n_\mathrm{cx,\HH}, \dot n_\mathrm{cx,\He}$), and the charge-exchange ionization rates (here defined as $\beta_\HH=\dot n_\mathrm{cx,\HH}/n_\p$, $ \beta_\He=\dot n_\mathrm{cx,\He}/(n_\HeI+n_\alp)$) of hydrogen and helium. 
    
	Clearly, in case \texttt{A} the only source in mass to the plasma mixture %[NOW IT IS SOURCE IN MASS TO THE PLASMA MIXTURE, BUT STILL NOT GOOD, AS I MENTIONED EARLIER. YOU PROBABLY MEAN THE CONTRIBUTION TO THE SOURCE TERM IN THE CONTINUIIY EQUATION.]] 
    comes from processes (\textit{b}) and (\textit{e})  involving the conversion of alpha particles into \HeI ions and \HI PUIs. Moreover,  the positive source  of \HI PUIs in the inner heliosheath (panel b) is only due to charge exchange of the fast, neutral SW atoms created in the supersonic SW. As expected, however, photoionization of \He atoms constitutes the major source of pickup \HeI ions in the supersonic SW. %[DO YOU MEAN THE ENTIRE HELIOSPHERE?] 
    As a consequence, the density of \HeI ions in case \texttt{B} becomes noticeably higher compared to case \texttt{A} (approximately by a factor of $\sim10$ in 
    %considering 
    the distant SW along the upwind direction). The same is true for their pressure, which becomes higher by a factor of $2.5\times10^2$. In turn, charge exchange of \HeI ions in the heliosheath becomes effective, causing helium ions to cool throughout the heliosphere and be converted into \He atoms (note the and negative net source term in the \HeI pressure equation). This produces \He ENAs, whose fluxes and energy spectra we are aimed to derive with this self-consistent model, and potentially observe with IMAP. In contrast, charge exchange involving alpha particles contributes to plasma heating within 200 AU of the Sun and to its cooling in the heliotail beyond 400 AU, where \alp ions are significantly hotter. This finding aligns with the result published in \cite{fraternale2024b}. Notably, we observe a non-negligible feedback of \HeI ions on the global solution, resulting in an increased heliosphere size.
    
    The boundaries of the kinetic grids are somewhat visible in Fig. \ref{fig:sources_NoPhi} and Fig. \ref{fig:sources_Phi}. These are statistical artifacts arising from particle splitting. As particles enter the finer grid domain and the splitting process is completed, the statistics improve. This effect is more pronounced in some panels, particularly for PUIs, because neutral SW atoms have very low densities and therefore poorer statistics. Although we did not address these artifacts in this study—this being the first simulation conducted with this model—we know that they can be easily mitigated by fine-tuning the particle-splitting and recombination parameters. Nonetheless, we have verified that these artifacts in the source terms do not affect the plasma solution. 
	A detailed analysis of all plasma components and properties of neutral atoms will be discussed in a separate paper, where we will also compare these results with those obtained assuming kappa-distributed plasma components.

	\section{Conclusions}\label{sec:conclusions}
We have extended our MHD/Kinetic model, implemented in MS-FLUKSS, to account for the presence of both \HeI and \alp ions in the solar wind, as well as \HI pickup ions, treating them as distinct fluids and in a self-consistent manner. The new model includes six charge-exchange processes and two photoionization processes, which are expected to be crucial for an accurate representation of neutral atom distributions throughout the heliosphere. This approach will be of importance for modeling and prediction of \HH and \He ENA fluxes to be observed by IMAP in the future, which is one of the objectives of our project. The functionality and performance of the new code have been successfully demonstrated.

	\begin{center}
		\begin{table}[h]
			\caption{\label{tab:performance} Performance comparison on 40 Haswell nodes of Frontera (56  cores and 192 GB memory per node), using 4 MPI tasks per node and 14 OpenMP threads per task. }
			%\footnotesize\rm
			\centering
            
			% \begin{tabular}{@{}*{3}{l}}
			% 	\br
			% 	& prev. model & current model \\ \hline
			% 	Total \# cells   & 308536410  &  308536410 \\    
			% 	nondimensional base-grid time step for plasma,  $\Delta t_p$  & 0.151  &  0.151 \\
			% 	plasma wall-time/step, $wt_p$ (s)  & 2.70  &  3.75\\
			% 	walltime, MPI\_Bcast, $wt_{\rm Bcast}$ (s)  & 4.95  & 7.0 \\ \hline
			% 	Kinetic runtime (physical time, yrs)  & 500   & 500  \\
			% 	Hydrogen \#particles  & 3.5e8   & 3.5e8  \\
			% 	Helium \#particles  & 2.2e8   & 2.2e8  \\
			% 	walltime, \HH sources calculation, $wt_{\rm S,H}$ (s)  & 321 & 365    \\
			% 	walltime, \He sources calculation, $wt_{\rm S,He}$ (s)  & 205 & 215     \\
			% 	max. virtual memory/node (MaxVMsize, GB)  & 89  & 123  \\
			% 	max RAM memory/node (MaxRSS, GB)  & 76  & 110  \\
			% 	\br
			% \end{tabular}
\begin{tabular}{@{}lcc@{}}
\br
\textbf{Description} & \textbf{Previous Model} & \textbf{Current Model} \\ \hline
Total number of cells, $N_\text{cells}$   & 308,536,410  & 308,536,410 \\    
Nondimensional base-grid time step for plasma, $\Delta t_p$  & 0.151  & 0.151 \\
Plasma wall-time per step, $wt_p$ (s)  & 2.70  & 3.75 \\
Wall-time for MPI\_Bcast, $wt_\text{Bcast}$ (s)  & 4.95  & 7.00 \\ \hline
Kinetic runtime (physical time, years)  & 500   & 500  \\
Number of hydrogen particles, $N_\text{H}$  & $3.5 \times 10^8$   & $3.5 \times 10^8$  \\
Number of helium particles, $N_\text{He}$  & $2.2 \times 10^8$   & $2.2 \times 10^8$  \\
Wall-time for hydrogen source calculation, $wt_\text{S,H}$ (s)  & 321 & 365 \\
Wall-time for helium source calculation, $wt_\text{S,He}$ (s)  & 205 & 215 \\
Maximum virtual memory per node, MaxVMsize (GB)  & 89  & 123  \\
Maximum RAM memory per node, MaxRSS (GB)  & 76  & 110  \\
\br
\end{tabular}            
		\end{table}
	\end{center}

	\section{Acknowledgements}
	This work is supported by NASA grants  80NSSC24K0267 and 80NSSC18K1649, and by the NSF-BSF grant 2010450. 
    The authors acknowledge the Texas Advanced Computing Center (TACC) at The University of Texas at Austin for providing HPC resources on Frontera supported by NSF award CISE-OAC-2031611, and also acknowledge the IBEX mission as part of NASA’s Explorer Program (80NSSC18K0237). 	Supercomputer time allocations were also provided by NASA High-End Computing Program award SMD-17-1537. FF and NP acknowledge support from the International Space Science Institute (ISSI) in Bern, through ISSI International Team project \#574 ``Shocks, Waves, Turbulence, and Suprathermal Electrons in the Very Local Interstellar Medium''.  We are  thankful to P. Swaczyna for the insightful discussions about the charge exchange processes.

	\section*{References}
	%\bibliography{bibliography_nov2024}

	\providecommand{\newblock}{}

\end{document}